\documentstyle[prl,twocolumn,aps]{revtex}
\begin{document}

\twocolumn[\hsize\textwidth\columnwidth\hsize\csname
@twocolumnfalse\endcsname
\title{Critical State Behaviour in a Low 
Dimensional Metal Induced by Strong Magnetic Fields}
\author{N.~Harrison}
\address{National High Magnetic Field Laboratory, LANL, MS-E536, Los
Alamos, New Mexico 87545}
\author{L.~Balicas}
\address{Instituto Venezolano de Investigaciones Cient\'{\i}ficas, 
Apartado 21827, Caracas 1020A, Venezuela}
\author{J.~S.~Brooks}
\address{National High Magnetic Field Laboratory, Florida State 
University, Tallahassee, Florida 32310}
\author{M.~Tokumoto}
\address{Electrotechnical Laboratory, Tsukuba, Ibaraki 305, Japan}
\date{\today}
\maketitle

\begin{abstract}
We present the results of magnetotransport and magnetic torque 
measurements on the $\alpha$-(BEDT-TTF)$_2$KHg(SCN)$_4$ 
charge-transfer salt within the high magnetic field phase,
in magnetic fields extending to 33~T and 
temperatures as low as 27~mK. While the high magnetic field phase 
(at fields greater than $\sim$~23~T) is expected, on 
theoretical grounds, to be either to a modulated charge-density 
wave phase or a charge/spin-density wave hybrid, the resistivity 
undergoes a dramatic drop below $\sim$~3~K within the high magnetic 
field phase, falling in an approximately exponential fashion at low 
temperatures, while the magnetic torque exhibits pronounced hysteresis 
effects. This hysteresis, which occurs over a broad range of 
fields, is both strongly temperature-dependent and has several of the 
behavioural characteristics predicted by critical-state models
used to describe the pinning of vortices in type II 
superconductors in strong magnetic fields. Thus, rather than 
exhibiting the usual behaviour expected for a density wave ground 
state, both the transport and the magnetic properties of  
$\alpha$-(BEDT-TTF)$_2$KHg(SCN)$_4$, at high magnetic fields,
closely resemble those of a type II superconductor.
\end{abstract}

]\narrowtext

\section{introduction}
Of all the quasi-two-dimensional (Q2D) organic charge-transfer 
salts that exist \cite{ishiguro1}, 
those of the composition $\alpha$-(BEDT-TTF)$_2M$Hg(SCN)$_4$, 
where BEDT-TTF stands for bis(ethylenedithio)tetrathiafulvalene and 
where $M=$~K, Tl, Rb or NH$_4$, have been, perhaps, the most difficult 
to understand \cite{wosnitza1,montpellier1}. While, the $M=$~NH$_4$ 
salt is a superconductor with a transition temperature  
$T_{\rm c}\sim$~1~K \cite{wang1}, 
the $M=$~K, Tl and Rb salts all undergo a transition into a more 
resistive state with a reconstructed Fermi surface below $T_{\rm 
p}\sim$~8 (in the $M=$~K and Tl salts) or 10~K (in the $M=$~Rb salt) 
\cite{wosnitza1,montpellier1}. 
The underlying physical reason for the behavioural 
differences between the isostructural $M=$~NH$_4$ and $M=$~K, Tl and 
Rb salts remains a contemporary issue. 
The $M=$~K salt has, nevertheless, been shown to 
become superconducting under uniaxial stress applied perpendicular to the 
conducting layers \cite{campos1,hirayama1}, and it has further been  
suggested that the $M=$~K and Rb salts could exhibit filamentary 
superconductivity even at ambient pressure 
\cite{ito0,brooks1,ito1}.

Unquestionably, the main physical traits of the $M=$~K, Tl and 
Rb salts at ambient pressure, at moderately low temperatures 
$T\lesssim T_{\rm p}$ 
and at magnetic fields $B<B_{\rm k}$, where 
$B_{\rm k}$ ($\approx$~23~T in the $M=$~K salt) 
is known as the kink transition 
field, are more typical of a density wave (DW) ground state
\cite{pratt1,kartsovnik1}. 
Yet, no direct evidence for a superlattice structure has been found. 
Most of the more recent experimental and theoretical surveys point
towards a charge-density wave (CDW) rather than a spin-density wave 
(SDW) being the more likely candidate
\cite{miyagawa1,mckenzie1,biskup1,harrison1}. 
The electrical transport is strongly magnetoresistive, with the increase 
in resistance below $T_{\rm p}$ becoming increasingly pronounced in a 
magnetic field \cite{pratt1,kartsovnik1,harrison4}. 
On the approach to $B_{\rm k}$, however, the magnetoresistance changes, 
becoming much lower above 
this field \cite{osada1}. The nature 
of the high magnetic field phase ($B>B_{\rm k}$), at low temperatures, 
has since been the subject of speculation 
\cite{mckenzie1,biskup1,house1,kartsovnik2,harrison2,hill1,honold1}. 
The only fully established fact is that the
transition field $B_{\rm k}$ is strongly first order, with 
hysteresis appearing both in the magnetoresistance and the 
magnetisation over a few Tesla interval in field.

From the theoretical perspective, $B_{\rm k}$ lies remarkably close to 
the critical field $B_{\rm c}$ at which one would expect a 
CDW, with a mere transition temperature of 8~K, to approach the 
Pauli paramagnetic limit \cite{mckenzie1,harrison1}. 
A first order transition could be expected, with the material transforming 
at higher fields either into a normal metal \cite{harrison1}, a
spatially modulated CDW \cite{mckenzie1} or a mixed CDW-SDW hybrid
\cite{zanchi1}. The latter, in particular, 
represents the CDW analogue of the Fulde-Ferrel phase that is
anticipated to occur in strongly type II superconducters, with 
suppressed orbital effects, over the equivalent region of the phase diagram
\cite{fulde1,larkin1}. Were either of these latter two phases actually to be 
realised at high magnetic fields, their physical properties have not 
yet been the subject of a thorough investigation. 
Experimentally, data has been published that could be considered 
conducive to the existence of a different thermodynamic phase above 
$B_{\rm k}$ at low temperatures \cite{kartsovnik2,kuhns1},
albeit that the magnetic data, at that time, was discussed largely in the 
context of SDW's with no metion of CDW's. Yet other data, 
concerning the unusual beheviour of the interlayer magnetotransport 
and induced currents in the magnetisation in pulsed magnetic fields, 
has been notionally connected with the quantum Hall effect (QHE) 
\cite{harrison2,hill1,honold1}. While a definitive experiment has not been 
performed that can firmly establish either of these latter two 
scenarios at high magnetic fields ($B>B_{\rm k}$), one cannot dispute 
the fact that this high magnetic field phase is somewhat exotic.

With a view to understanding more about the high magnetic field 
phase (at fields above $B_{\rm k}$), in this paper we 
present the results of extensive measurements of the interlayer magnetotransport 
and the magnetic torque on several $\alpha$-(BEDT-TTF)$_2$KHg(SCN)$_4$
crystals over a broad range of temperatures, 20~mK~$<T<$~10~K, and in 
continuous magnetic fields of up to 33~T. The observed behaviour of 
the magnetotransport and magnetic torque is neither typical of that 
expected for a DW system, nor does it support recent claims of the QHE in 
this material. For example; as the sample is cooled in a magnetic field, 
the resistivity undegoes an abrupt drop at $\sim$~3~K, at all fields 
$B\gtrsim B_{\rm k}$, with the drop being particularly pronounced at 
integer Landau level filling factors, falling exponentially by roughly two 
orders of magnitude. The occurrence of a change in slope in the magnetic 
torque at precisely the same temperature, on field-cooling the sample, 
lends further 
support to the existence of a different low temperature thermodynamic 
phase below $T_{\rm c}\lesssim$~3~K \cite{kartsovnik2}. 
At lower temperatures still,
({\it i.e.} $T\lesssim$~2~K), the field-dependence of the magnetic torque 
develops a pronounced hysteresis at all fields above $B_{\rm k}$ that increases 
approximately exponentially with decreasing $T$, but with the 
hysteresis being particularly strong at half integer Landau level 
filling factors. Only at much lower temperatures ($T\sim$~27~mK) does the 
hysteresis become more pronounced at integer filling factors 
\cite{harrison2}. While 
the hysteresis throughout the entire high magnetic field phase is 
difficult to explain in terms of conventional 
DW ground states or the QHE, it has many of the features 
of a critical state model like those used to explain magnetic hysteresis in 
type II superconductors. In the absence of any other model specific 
to the high magnetic field phase of 
$\alpha$-(BEDT-TTF)$_2$KHg(SCN)$_4$,
the irreversible hysteretis, together with the sharp drop 
in the resistivity at low temperatures, is remarkably similar to the
behaviour of superconductor in strong magnetic fields. Such behaviour 
cannot be explained consistently in terms of the QHE or any 
conventional DW ground state.
\section{experimental}
The samples chosen for this study, referred to hereon as 1, 2 and 3, 
were grown using conventional 
electrochemical means \cite{tokumoto1}, while static magnetic fields 
extending to $\sim$~33~T were provided by the 
the National High Magnetic Field Laboratory, Tallahassee. 
Temperatures between $\sim$~20~mK and $\sim$~1.6~K were provided 
by a top-loading dilution refrigerator, with higher temperatures 
obtained using a $^3$He refrigerator. 
Magnetotransport measurements 
were made using standard four wire techniques with low frequency 
($\sim$~10~Hz) currents of
between 1 and 10$\mu$A applied perpendicular to the 
conducting layers. For the magnetic torque measurements,
the samples were mounted on the moving plate of a phosphor bronze
capacitance cantilever torque magnetometer, 
which was itself attached to a rigid but rotatable
platform in such a way that the axes of torque and rotation were parallel 
to 
each other, but perpendicular to the applied magnetic field $B$.
The angle between $B$ and the normal to the capacitance 
plates was approximately the same as the angle $\theta$ between $B$ 
and the normal to the conducting planes of the sample.
The capacitance of $\sim$~1.3~pF was measured by means of a capacitance 
bridge energized at 5~KHz and 30~V~(rms); the largest change in C 
observed throughout the experiment at the highest magnetic field was 
$\sim$~0.04~pF, corresponding to a net angular displacement of 
$\sim$~0.1$^\circ$. Torque interaction effects at $\theta =$~7$^\circ$ 
were therefore not a significant factor. To eliminate possible 
artefacts due to time constants of the instrumentation setup or the data 
acquisition system, the capacitance was measured with the time constant 
of the lock-in amplifier set to a 
low value of 10~ms. Furthermore, both the output of the lock-in amplifier 
and the shunt voltage that determines the current flowing in the Bitter 
magnet coils, were measured using a digitizer with resolution of 
$\sim$~50~$\mu$s, while the field was swept slowly at $\sim$~8~mTs$^{-1}$.
\section{temperature-dependence of the magnetoresistance}
Examples of the interlayer magnetoresistance measured in 
$\alpha$-(BEDT-TTF)$_2$KHg(SCN)$_4$ sample 1 at selected temperatures 
are shown in Fig. \ref{magres}. The magnetoresistance displays the 
usual 
behaviour observed for this material, reaching a maximum at 
$\sim$~10~T before falling again in a linear fashion as the kink transition 
field is approached \cite{pratt1,tokumoto1}. In accordance with these
earlier studies, the 
oscillations within the low magnetic field DW phase exhibit a 
pronounced second harmonic.
Above the kink transition, at temperatures higher than $\sim$~3~K, 
the oscillations grow rapidly in amplitude with increasing field. The 
behaviour of the interlayer Shubnikov-de Haas (SdH) waveform
over this range of temperatures is well understood in these materials 
\cite{harrison3,datars1}.
At integer Landau level filling factors, when the chemical potential 
$\mu$ is situated in a Landau gap, the SdH maxima
increase with decreasing temperature in a insulating-like fashion.
Conversely, at half-integer filling factors, the resistivity of the minima 
behave in a metallic fashion. 
This is entirely consistent with the theoretically expected behaviour 
of a quasi-two-dimensional metal in a magnetic field 
\cite{harrison3,datars1}. 

What is not predicted by the simple theory 
is the abrupt inversion of the waveform at low temperatures, 
resulting in 
the high temperature ($T\gtrsim$~3~K) SdH maxima becoming minima at 
low temperatures ($T\lesssim$~3~K) with a resistivity several times 
lower than that of the sample at zero field. Figure
\ref{magres} does not constitute the first observation of this 
effect; previous studies had reported this effect in both 
continuous and pulsed magnetic fields \cite{hill1,honold1}. Because no
phase inversion is observed in the dHvA effect \cite{hill1,honold1}, 
which is entirely a thermodynamic function of state, this
implies that the Landau level structure remains largely unchanged 
over the same temperature range. It was the occurrence of this phase inversion 
only in the magnetotransport that was then attributed to the effect 
of a chiral Fermi liquid \cite{hill1,honold1}, following a suggestion 
that the QHE may be taking place at high magnetic fields in this 
material \cite{harrison2}. Because the interlayer resistance of the bulk of the 
sample becomes insulator-like at integer filling factors according to 
magnetotransport theory \cite{harrison3,datars1}, higher conductivity 
chiral Fermi liquid states were proposed to take over the majority of 
the interlayer conductance. For this conjecture to explain the most recent 
data ({\it i.e.} Fig. \ref{magres}), the interlayer conductivity of 
the edge states, which occupy $\sim$~1 part in 10$^4$ of the sample 
cross section, would have to be both strongly temperature-dependent 
and attain a conductivity at least 10$^6$ times higher than that of the 
field-averaged (or background) conductivity of the bulk.

Since this conjecture was made, the interlayer conductance of chiral 
surface states has 
been measured directly in semiconductor superlattices that at the same 
time quite clearly exhibit the QHE \cite{druist1,zhang1}.
Not only is the conductance of these states found to be 
temperature-independent, 
as predicted by chiral Fermi liquid theory \cite{chalker1,balents1},
but their net conductance is also observed not to be particularly high, resulting 
in only a weak suppression of the SdH maxima. This behaviour is
therefore quite different from that observed in 
$\alpha$-(BEDT-TTF)$_2$KHg(SCN)$_4$. When coupled with the fact that 
quantized Hall plateaux proportional to $\hbar/ie^2$ have not been 
observed \cite{harrison4} (in the $M=$~Tl salt), the  
chiral Fermi liquid model can no longer be considered valid.

A more distinct behaviour emerges when the resistivity, 
both at integer 
and half-integer filling factors, is plotted versus temperature in 
Fig. \ref{res}. For two different samples, the resistivity varies 
weakly with temperature for $T\gtrsim$~3~K, but then undergoes an 
abrupt drop for $T\lesssim$~3~K. 
This drop is particularly pronounced at integer filling 
factors ({\it i.e.} when $\mu$ is situated in the 
Landau gap of the Q2D pockets), but is also clearly discernable at 
half-integer filling factors ({\it i.e.} when $\mu$ is situated in 
the middle of a Landau level). This behaviour is somewhat similar to that 
observed by Kartsovnik {\it et al.} \cite{kartsovnik2}, except that 
in the present case, at integer filling factors in sample 1, the resistivity is 
observed to fall by two orders of magnitude between 3~K and 490~mK, 
reaching 1.6~$\Omega$ at the lowest temperature. 
While 1.6~$\Omega$ does not sound 
that low, it is $\sim$~2000 times lower than the room temperature 
resistance of 3.6~k$\Omega$ for this sample, and this is even in the 
presence of a $\sim$~30~T magnetic field.
In spite of the 
fact that the samples were cooled in a dilution refrigerator on a subsequent 
experimental run to temperatures as low as 20~mK, zero resistivity 
was never observed. 
The abrupt change in resistivity versus temperature $\rho(T)$ has 
previously
been interpreted as evidence for a phase transition \cite{kartsovnik2}, and 
the present study appears to support this claim. An abrupt drop in the 
resistance, as observed here, cannot be explained by chiral Fermi 
liquid theory \cite{chalker1,balents1}, nor is the occurrence of a transition 
at half-integer filling factors (as well as integer filling factors) compatible 
with the QHE. Thus, it would appear to be the case that the inversion of 
the SdH waveform is
merely an artefact of the drop in resistivity being more pronounced at 
integer filling factors.

From a phenomenalogical point of view, the low temperature 
interlayer resistivity in Fig. \ref{res} obeys an approximate
$\rho\propto\exp(T/T_0)$ law, with $T_0\sim$~520~mK at integer filling 
factors for both samples 1 and 2. Such an exponential law is clearly 
unlike that of a normal metal, and the very observation of a 
sudden drop in resistivity is not expected for a DW ground state. 
Transitions into DW ground states are invariably followed by an 
immediate increase in resistivity on decreasing the temperature 
through the transition \cite{gruner1}, owing to a net loss of 
carriers. 
A drop in resistivity, of the form observed here, more 
commonly precedes a superconducting state. One could argue that,
while a zero resistivity state is not observed, zero resistance states 
are always more 
difficult to come by in strong magnetic fields owing to vortex dynamics
\cite{tinkham1,hightcconf}. It is also
interesting to note that the exponential dependence of the resistivity 
at low temperatures is strikingly similar to that observed in the 
filamentary superconductor La$_2$CuO$_{4-y}$ \cite{hsu1}.
\section{temperature-dependence of the magnetic torque}
The temperature-dependence of the magnetotransport alone 
cannot be considered as proof for a transition into a new 
thermodynamic phase. This usually requires measurement of a 
thermodynamic function of state, such as specific heat or 
magnetisation. Specific heat measurements have thus far confirmed the 
existence of a second order transition into a low temperature phase 
within the low magnetic field regime ($B<B_{\rm k}$) \cite{kovalev1}.
Within this same phase, magnetic moment measurements, made using a 
superconducting quantum interference device (SQUID) in fields of 5~T, 
have shown that the susceptibility drops most notably when the field is 
oriented within the conducting planes \cite{sasaki1}. Such a behaviour 
could be expected either for a CDW or a SDW phase following a net loss in 
Pauli paramagnetism with the opening of gaps on the Fermi surface 
\cite{gruner1}. Only a very weak change, however, was observed when the 
magnetic field was oriented perpendicular to the conducting layers. 
Sasaki {\it et al.} notionally connected this anisotropy with 
antiferromagnetism associated with a SDW ground state \cite{sasaki1}, 
although there has since not been any compelling evidence 
supporting this hypothesis \cite{miyagawa1,pratt2}. 

An alternative, yet rather trivial, explanation could be that when the field is 
oriented perpendicular to the conducting layers,
the loss in Pauli paramagnetism is balanced by a loss in Landau 
diamagnetism following the partial opening of gaps on the Q2D 
Fermi surface pockets due to a CDW. Gaps on the Q2D pocket are, after 
all, predicted by {\it all} models of the reconstructed Fermi surface 
\cite{harrison0}. Perhaps the strongest evidence for the latter 
explanation is that the magnetic susceptibility is completely 
isotropic for fields oriented within the planes, as demonstrated by
the magnetic torque measurements of Christ {\it et al.} \cite{christ1}. 
This detail would be difficult to explain in terms of antiferromagnetism, 
but is quite easy to explain in terms of Landau diamagnetism, which 
in a Q2D conductor, manifests itself only perpendicular to the 
conducting layers.

The essential advantage of magnetic torque measurements is that they are 
sensitive only to the magnetic anisotropy, with the 
net torque being given by the product
\begin{equation}\label{cross}
    \mathbf{\tau}=\mathbf{M}\times\mathbf{B}.
\end{equation}    
Thus, if we consider the susceptibility to be resolved into its components 
perpendicular to the conducting layers $\chi_\bot\equiv M_\bot/B$ and parallel 
to the conducting 
layers $\chi_\|\equiv M_\|/B$, the net magnetic torque is then given by
\begin{equation}\label{torque}
    \tau=\frac{1}{2}\big[\chi_\| -\chi_\bot\big]B^2\sin{2\theta},
\end{equation}
where $\theta$ is the angle between the magnetic field and the normal 
to the conducting layers ({\it i.e.} the ${\bf b}$ axis of the crystal). 
With the first term of Equation (\ref{torque}) being dominant and negative, 
according to Sasaki {\it et al}. \cite{sasaki1},
the change in anisotropy on entering the low temperature DW phase 
exerts a net negative torque on the sample, that attempts to 
align the ${\bf b}$ axis of the sample more closely with $B$.
This is exactly the signature observed by Christ {\it et al.} 
\cite{christ1}. 
For a constant anisotropy $\chi_\|-\chi_\bot$, the negative torque 
should increase in a manner that is proportional to $B^2$. The fact 
that this is not the case, as shown originally by Christ {\it et 
al.} \cite{christ2} and again in Fig. \ref{torquevsB}, could be 
considered as further supportive evidence for our explanation in terms of Landau 
diamagnetism, whereby the Landau diamagnetic contribution from the Q2D 
pocket returns gradually at higher magnetic fields due to magnetic 
breakdown across the CDW gaps.

Even though the anisotropy becomes reduced at high magnetic fields, 
a change in magnetic torque on field-cooling the sample is still 
visible in Fig \ref{torquevsT}. This is perhaps 
assisted by the increasing sensitivity ($\propto B^2$) of torque 
magnetometry 
with field. While the reason for the continued anisotropy 
at high magnetic fields is not obvious, the abrupt change in slope in fields as 
high as $\sim$~30.1~T clearly indicates the continuing presence of a 
thermodynamic phase boundary. 
The field of 30.1~T was chosen because, at this field, 
$\mu$ is exactly that for a half-integer filling factor, thereby eliminating 
any possible contribution to the torque from the dHvA signal 
\cite{kartsovnik2}. 
If our interpretation of the magnetic measurements of Christ 
{\it et al.} and Sasaki {\it et al.} is correct, this could implies that the 
loss of Landau 
diamagnetism on entering the low temperature, high magnetic field phase 
continues to be the dominant factor. Were part of the sample to 
become superconducting at high magnetic fields, as might be suggested 
from the behaviour of the resistivity,
it is unclear how this would affect the results. 
Usually, the pure Abrikosov diamagnetism of the vortex 
lattice of a Q2D superconductor is largest when the field is 
applied perpendicular to the conducting layers \cite{tinkham1}. 
The torque resulting 
from the net gain in Abrikosov diamagnetism 
should therefore have the opposite sign of that coming from the loss 
in Landau diamagnetism. For our results to be consistent with 
the establishment of a high magnetic field-induced superconducting 
state, the Landau diamagnetism would have to be larger in magnitude 
than the Abrikosov diamagnetism at these fields. Whether this is the 
case depends on parameters which are presently unknown.
\section{hysteresis in the magnetic torque}
One of the most intriquing apsects of this material 
at high magnetic fields is its magnetic hysteresis. 
An example of the 
magnetic torque of sample 3 measured both on rising and falling 
magnetic fields at 27~mK is shown in Fig. \ref{hystvsB}. Apart from 
the fact that the temperature is lower in the current work, the 
hysteresis is very similar to that observed by Christ {\it et al.} 
\cite{christ3}, although this effect was largely disregarded 
as originating from a ``complex magnetic 
groundstate,'' with the original interpretation of the anisotropic susceptibility 
work of  Sasaki {\it et al.}\cite{sasaki1} in terms of antiferromagnetism, 
essentially forming the foundation of their argument.

Consideration of the fact that only the anisotropy of the 
magnetic suscptibility gives rise to magnetic torque is the key to 
understanding the behaviour of the magnetic torque in this 
material. Hence, 
because the dHvA effect is the oscillatory component of 
the Landau diamagnetism, thereby only involving orbital effects 
within the conducting planes, it manifests itself as large oscillations 
of the magnetic torque that are rather straightforward to interpret. 
At low magnetic fields ($B<B_{\rm k}$), the oscillations exhibit 
the usual double-peaked structure that has been interpreted both 
as spin-splitting \cite{pratt1,tokumoto1,uji1,sasaki2,sasaki3,sasaki4}
and as the frequency-doubling effect that accompanies a pinned CDW or 
SDW phase \cite{harrison1}. This is well known to be an effect 
observed at small angles ($\theta\lesssim$~20$^\circ$) in this 
material (as well as the $M=$~Tl and Rb salts), which occurs only 
within the low magnetic field DW phase. On increasing the field
at these low temperatures ({\it i.e.} 
$T=$~27~mK), the kink transition is observed to be extremely abrupt, with a 
small but reproducible spike feature signalling the transformation into the
high magnetic field regime at $B\sim$~24.2~T in sample 3. 
Above this field, the dHvA oscillations develop the characteristic
triangular form that has been observed within the high magnetic field 
phase \cite{christ2,christ3,uji1}. 

The sudden change in the waveform that occurs at $\sim$~24.2~T on the 
rising magnetic field and at $\sim$~21.4~T on the falling magnetic 
field is indicative of a first order phase transition between 
two distinctly different regimes. Because 
the magnetic torque is entirely reversible within the low 
magnetic field DW phase, provided the field does not exceed $\sim$~24.2~T,
both the dHvA oscillations and the monotonic background torque remain
unchanged when the direction of sweep of the magnetic field is reversed.
Similarly, 
when the field sweep direction is reversed within the high magnetic 
field phase, the dHvA oscillations continue to have the same 
triangular form, provided that the field is not swept below 
$\sim$~21.4~T. This type of behaviour indicates that the low and high 
magnetic field phases are immiscible, with the formation 
of domains being energetically unfavourable. Because it costs energy 
to mix the two phases, the low magnetic field DW phase is 
``supercooled'' on increasing the magnetic field until the free 
energy difference between the two phases is no longer sustainable.
To help understand this behaviour in 
$\alpha$-(BEDT-TTF)$_2$KHg(SCN)$_4$,
it is instructive to consider a simplified model where the 
zero temperature free energies of the low and high magnetic 
field regimes are approximated as
\begin{equation}\label{freeCDW}
    F_0\approx -g_{\rm 1D}\bigg[\frac{\Delta_0^2}{2}
    -h^2\bigg]
\end{equation}
and 
\begin{equation}\label{freex}
    F_x\approx -g_{\rm 1D}\frac{\Delta_x^2}{2}
\end{equation}
respectively, with $g_{\rm 1D}$ being the density of Q1D states and 
$h=g\mu_{\rm B}B/2$ \cite{mckenzie1,harrison1,zanchi1}, and 
we assume, to first order, that the high magnetic 
field phase is not affected by magnetic field. Here, the 
subscript $0$ denotes the CDW phase that is stable at low magnetic 
fields, while $x$ denotes the spatially modulated or CDW-SDW hybrid 
phase \cite{mckenzie1,zanchi1}. If we then assume the 
free energy of the domain term $F_{\rm mix}\approx\epsilon y(1-y)$ to be 
approximately parabolic ({\it i.e.} the lowest order even funtion of $y$), 
with $y$ being the fraction of the material in 
the low magnetic field phase, then no domain structure can ever be stable 
provided $\epsilon\geq 0$. Since the total free energy is 
\begin{equation}\label{mix}
    F_{\rm tot}\approx yF_0+(1-y)F_x+\epsilon y(1-y),
\end{equation}
it then follows that, on increasing the magnetic field within the low 
magnetic field regime, within which $y=1$, the material 
cannot ``snap'' into the high magnetic field phase until $\partial 
F_{\rm tot}/\partial B\lesssim 0$. Conversely, on 
decreasing the magnetic field from within the high magnetic field 
regime, within which $y=0$, 
the material cannot snap into the low magnetic field phase 
until $\partial F_{\rm tot}/\partial B\gtrsim 0$. The kink 
transitions for rising and falling magnetic field are therefore 
\begin{equation}\label{kinks}
    h_{\rm k}=\sqrt{\frac{\Delta^2_0-\Delta^2_x}{2}\pm
    \frac{\epsilon}{g_{\rm 1D}}},
\end{equation}
with the $\pm$ sign corresponding to the direction of sweep of the 
magnetic field. The abruptness of the transitions should be accompanied 
by the release of latent heat, although this cannot be detected in 
the present isothermal experiment.
If we assume the strength of the coupling to be similar in the two 
ordered 
regimes, then the ratio of order parameters $\Delta_x/\Delta_0$ is equal 
to the ratio of transition temperatures $T_{\rm c}/T_{\rm p}\sim 3/8$, 
and we obtain $\Delta_0\sim$~4.1~meV and $\Delta_x\sim$~1.5~meV on 
inserting $B_{\rm c}\equiv B_{\rm k}\sim$~23~T. Meanwhile, for $1/g_{\rm 
1D}\sim$~30~meV, we obtain $\epsilon\sim$~24~$\mu$eV.

The hysteresis that 
starts at magnetic fields above $\sim$~24.5 has quite a different form 
from that associated with the kink transition.
Since (1) the dHvA oscillations have the same shape and 
size between rising and falling magnetic fields and (2) the hysteresis 
takes place continuously over a wide interval in field (from 
$\sim$~24.5~T up to the highest available field of $\sim$~33~T), this new 
type of hysteresis appears not to be a consequence of a second prolonged 
first order phase transition as a function of magnetic field, as was 
suggested by McKenzie \cite{mckenzie1}. For a first order 
phase transition to take place over such an extended interval of 
field, either (1) the free energies of the two competing phases would 
have to be very similar but with slightly different dependences on $B$ or $\mu$, 
or (2) the mixing term $\epsilon$ would 
have to be very large (and negative). Were this the case,  
the hysteresis would have to result from two competing CDW or CDW-SDW 
hybrid phases with different anisotropic magnetic susceptibilities, 
different nesting vectors and therefore different reconstructed Fermi surface 
topologies. Any difference in Fermi surface topology between rising 
and falling magnetic field should ultimately 
manifest itself on the dHvA oscillations. This 
is to be expected since the peak-to-peak amplitude of the dHvA waveform 
is directly proportional to the number of Q2D states, while the gradient 
on the falling side of the oscillation ({\it i.e.} at integer filling) 
is proportional to the 
density of background states \cite{harrison3}. As it turns out, 
however, there is no detectable change in the peak-to-peak height nor 
in the falling slope of the dHvA oscillations between rising and 
falling fields in Fig. \ref{hystvsB}, at least for $B\gtrsim$~26~T. 
In fact, from the fraction of the waveform 
$\gamma=g_{\rm 2D}/(g_{\rm 2D}+g_{\rm 1D})$ over which 
the dHvA magnetisation increases with field \cite{harrison3}, it is 
rather straight forward to infer 
that the density of background states is 68~$\pm$~5~\% that of the Q2D 
pocket, both on rising and falling fields, therefore having the same 
density of states, within experimental uncertaintly, as the Q1D Fermi surface 
sheets prior to nesting \cite{harrison0}. 
This could imply that the Q1D states are not nested (or ``ungapped''). 
Indeed, it has been a common
conclusion of {\it all}
quantum oscillation and angle-dependent magnetoresistance 
oscillation measurements, that the Fermi surface 
appears to be unreconstructed within the high magnetic field phase 
\cite{pratt1,house1,harrison3}. Alternatively, the high magnetic 
field phase could be a superconductor, whereby (in the absence of a
${\bf Q}$-vector) the gap automatically remains pinned to 
the 
oscillating chemical potential, enabling the Q1D Fermi surface sheet 
to continue affecting the waveform of the dHvA oscillations. One 
could also
envisage a scenario whereby the ${\bf Q}$ vectors of a CDW or
CDW/SDW phase adjust themselves 
continuously in order to maintain the gap symmetrical with respect to the 
oscillating chemical potential; this situation would therefore contrast 
from that in the low magnetic field phase of 
$\alpha$-(BEDT-TTF)$_2$KHg(SCN)$_4$ \cite{harrison1} and that in 
NbSe$_3$ \cite{harrison4}. 
In either case, the size and shape of the quantum oscillations indicates 
that there is no discernable difference in the density of states between rising 
and falling magnetic fields, therefore making it somewhat unlikely 
that the electronic ground states could be different between rising and 
falling magnetic fields. The fact that no hysteresis is observed in 
the magnetotransport further implies that it cannot be the electronic 
structure which is itself hysteretic.

The only other possibility is that the hysteresis originates from some 
form of dynamic magnetism involving 
the ferromagnetic alignment of spins or circulating currents. 
Whenever magnetic hysteresis occurs, whether it originates from 
ferromagnetism, metamagnetism, vortex pinning in a type II 
superconductor or even induced currents in a quantum Hall system, the 
tendency is always to retain magnetic flux within the sample. Using 
the field-cooled magnetisation as a point of reference (or the 
magnetisation averaged between up and down sweeps), the magnetisation is 
always slightly more diamagnetic on the rising field and slightly 
more paramagnetic on the falling field. Only the 
anisotropy of the magnetic susceptibility gives rise to magnetic 
torque: the sign of the hysteresis observed here thus 
implies that it involves magnetic 
moments that are predominantly oriented perperdicular to the conducting 
planes, or equivalently, currents flowing within the conducting planes.
Clearly, $\alpha$-(BEDT-TTF)$_2$KHg(SCN)$_4$ contains no 
magnetic ions with partially filled $d$- or $f$-electron shells, so 
we can eliminate this rather trivial source of ferromagnetism or
metamagnetism. The hysteresis in Fig. \ref{hystvsB} is therefore more 
easy to explain in terms of induced currents. 
The fact that the hysteresis is observed only in the magnetic torque 
and not in the magnetotransport could be considered consistent with this 
hypothesis.
These cannot be quantum Hall currents, however,
since the hysteresis is observed at all fields 
$B\gtrsim$~24.5~T, both at integer and half-integer filling factors. 
\section{irreversible processes}
The behaviour that is probably most supportive of the notion of 
induced persistent currents, is the  
irreversibility of the magnetic torque on 
stopping or reversing the sweep direction of the magnetic field. 
In most metals, including those with DW phases, the magnetisation is 
usually reversible, as is the case, for example, within the 
low magnetic field DW phase ({\it i.e.} see Fig. \ref{hystvsB}).
On sweeping the field back and forth within the high magnetic 
field phase, on the other hand, even over a very small interval in field, 
the magnetic torque of 
$\alpha$-(BEDT-TTF)$_2$KHg(SCN)$_4$ is observed 
to be entirely irreversible, 
giving rise to a hysteresis loop of the form shown in Fig. 
\ref{loop}(a). In fact, the nature of the hysteresis has several of the 
characteristic features of a critical state model, such as the Bean 
model \cite{tinkham1,bean1} that is used to describe 
magnetic hysteresis caused by vortex 
pinning in type II superconductors. This becomes particularly evident 
in Fig. \ref{loop}(b) when the monotonic background and dHvA 
contributions are subtracted; the background is obtained by 
averaging data taken on full up and down sweeps of the magnetic field. 
In Fig. 
\ref{loop}(b) we can see that, on reversing the sweep direction of the magnetic 
field, the magnetic hysteresis increases gradually
until a critical value is reached. In type II superconductors, this 
would be proportional to the critical current density $j_{\rm c}$. 
In Fig. \ref{loop}(b), the induced magnetisation corresponding to the 
critical state is 
that obtained by performing full up and down sweeps [also shown in 
Fig. \ref{loop}(b)]. By ``critical state'' we imply that the 
sample has a tendancy to trap flux exactly like a type II superconductor, 
with the local currents determining the local magnetic field gradient. 
While the unit of trapped flux in type II superconductors is the 
vortex which contains a net flux equivalent to that of the flux 
quantum, it is unclear what the equivalent unit of trapped flux is in 
the high magnetic field phase of $\alpha$-(BEDT-TTF)$_2$KHg(SCN)$_4$.

It nevertheless follows from the critical state model as normally 
applied to type II 
superconductors, that as soon as the direction of 
sweep of the magnetic field is reversed, the initial slope of the magnetisation 
with respect to magnetic field is given by
\begin{equation}\label{derivative}
    \chi=\frac{\partial M}{\partial B}=-\frac{f[1-\eta]}{\mu_0}, 
\end{equation}
where $-1/\mu_0$ corresponds to perfect diamagnetism,
$\eta$ is the demagnetisation factor and $f$ is the volume fraction of 
the sample in which pinning occurs.
This would then be, perhaps, as close 
to performing a Meissner-type experiment as one could get in extreme 
magnetic fields. If we consider that the anisotropic hysteresis originates 
from currents flowing only within the conducting planes and that
$\eta\sim$~0.5 (as for a cylinder), upon taking the initial slope of the 
magnetic torque in Fig. \ref{loop}(b), 
we obtain $f\sim$~1~\% for sample 3 (of
volume $\sim$~0.8~mm$^3$). Thus, were the hysteresis in Fig. \ref{loop} 
really caused by vortex pinning, this would imply either that only 1~\% of the 
vortices are pinned, or that only 1~\% of the sample is 
superconducting. This conclusion, that only part if the sample is 
pinned or superconducting, could be considered consistent 
with the failure of the sample to transform into a bulk zero 
resistance state. The value of $f\sim$~1~\% would, of course, 
be underestimated were any of the supercurrent to flow perpendicular to 
the conducting planes via Josephson tunneling. The value of 1~\% 
should therefore be considered as a lower limit.
That it is necessary to sweep the magnetic field by 
$\sim$~0.4~T before the trapped flux in the sample is completely reversed, could 
either imply that the local field difference in parts of the sample reaches 
strengths of order 0.4~T, or that shape effects, which give rise to 
demagnetisation, are important.

Another property of the critical state model as applied to 
type II superconductors, is the that they often 
give rise to relaxation phenomena, through thermally assisted flux 
flow or quantum flux creep \cite{tinkham1,hightcconf,highBsup}.
One way to test whether such effects occur in 
$\alpha$-(BEDT-TTF)$_2$KHg(SCN)$_4$ is to monitor the dependence of 
the width of the hysteresis loop on sweep rate. Non-linear 
current-voltage characteristics have already been detected in pulsed magnetic 
fields experiments, although it is quite likely, should
 $\alpha$-(BEDT-TTF)$_2$KHg(SCN)$_4$
(or the $M=$~Tl salt) be a superconductor, that these experiments 
were performed close to the flux flow regime 
\cite{harrison2,honold2}. On sweeping the field at rates
between 8~mTs$^{-1}$ and 50~mTs$^{-1}$ in Fig. 
\ref{loop}, the hysteresis increases by no more than 
10~\%, implying that the relaxation rate is rather low.
A more notable degree of relaxation of the magnetic torque is 
observed by stopping the sweep of the magnetic field abruptly and then 
observing changes over several minutes. At integer filling factors 
[Fig. \ref{relax}(a)], the magnetic torque 
decays much more rapidly than at half-integer filling factors [Fig. 
\ref{relax}(b)] and is considerably more noisy. On fitting the Anderson-Kim 
flux-creep model \cite{tinkham1} (making the substitution of $\tau$ for $M$)
\begin{equation}\label{Anderson}
    \tau_{\rm in}\approx\tau_{\rm crit}
    \bigg[1-\frac{\alpha_1}{2\alpha_{\rm c}}\ln{t}\bigg],
\end{equation}
which is thought to be applicable at low 
temperatures in most type II supeconductors, 
the better fit is obtained at half-integer filling 
factors, with the characteristic parameter $\alpha_1/\alpha_{\rm c}$ 
for logarithmic decay being of order 1/620. 
This is certainly within an order of magnitude of the theoretical 
prediction for most type II superconductors \cite{tinkham1}.
\section{field and temperature-dependence of the hysteresis}
Perhaps, a further analogy can be made with type II superconductors on 
consideration of the temperature dependence of the width of 
the hysteresis loop. In many type II superconductors, this is found to be 
strongly dependent on temperature, 
caused, for example, by the dependence of the vortex 
pinning potential $U_{\rm c}(T)$ on $T$ \cite{hightcconf}. 
Such a behaviour is also 
observable for the critical state of 
$\alpha$-(BEDT-TTF)$_2$KHg(SCN)$_4$, as shown in Fig. 
\ref{hystvsT}(a), albeit the hysteresis contains an oscillatory 
component. While no hysteresis could be detected at $T\sim$~4~K, it 
is already quite pronounced by 1.16~K, 
particularly at half-integer filling factors. Half-integer filling 
factors, realised whenever $F/B+1/2$ assumes an integer value,
are depicted in Fig. \ref{hystvsT} by solid vertical lines. Note that the 
oscillatory component of the width 
of the magnetic hysteresis loop oscillates in quadrature with the dHvA 
oscillations [extracted in Fig. \ref{hystvsT}(b)], but in phase with 
the density of states, with the hysteresis and the density of states at 
$\mu$ both exhibiting maxima at half-integer filling factors. Clearly,
as discussed in the preceding section,
it cannot be the dHvA oscillations themselves that are hysteretic
because (1)
the dHvA and oscillatory component of the hysteresis effects are at 
quadrature with respect to each other and (2) the 
actual dHvA amplitude is only relatively weakly 
dependent on temperature over the same range 27~mK~$<T<$~1.15~K, 
shown in Fig. \ref{hystvsT}(b). This together with the absence of any 
hysteresis in the SdH waveform, again, is consistent with our 
hypothesis in terms of induced currents.

As $T\rightarrow 0$, the 
hysteresis becomes double-peaked, with the second sharper and more 
strongly temperature-dependent peak emerging at integer filling factors. 
Sharp 
peaks in the hysteresis were observed at integer filling factors in 
pulsed magnetic fields \cite{harrison2,honold1}, 
and it was this ``coincidence'' that 
led to the data being interpreted in terms of the QHE.
When plotted as a function of temperature in Fig. \ref{exphyst}, the 
hysteresis is strongly dependent on temperature both at integer and 
half-integer filling factors, with the form of the $T$-dependence not being too 
different from that typically observed in high temperature superconductors 
\cite{hightcconf}.
\section{phase diagram}
It is instructive to construct a 
tentative phase diagram for $\alpha$-(BEDT-TTF)$_2$KHg(SCN)$_4$, from 
the data accumulated, at least for 
fields applied within $\theta\sim$~10$^\circ$ of the ${\bf b}$ axis of
the crystals.
Solid symbols have been chosen for the thermodynamic data, in 
Fig. \ref{diagram}, which is 
often the most reliable indicator for a thermodynamic phase 
transition. Solid squares denote the change in slope of the field-cooled 
magnetic torque extracted from Fig. \ref{torquevsT}. At high magnetic 
fields, this agrees well with the resistive transition at low temperatures
(indicated by open circles), as, for example, 
can be seen in both samples 1 and 2 in Fig. \ref{res}. Note 
that this transition is not sample-dependent; only the extent to 
which the resistivity decreases at low temperatures is sample-dependent, 
varying somewhat between samples 1 and 2.

At low magnetic fields, the field-cooled magnetic torque data agrees 
rather well 
with the transition into the low magnetic field DW phase obtained from 
specific heat measurements \cite{kovalev1}, represented here by solid 
triangles.
In the vicinity of the kink transition, however, the second order 
phase boundary, on cooling, becomes more difficult to extract from 
resistivity data, owing to the competition between different types of 
resistivity behaviour from the low magnetic field DW and exotic high 
magnetic field phases. For this reason, these points, depicted here as open 
squares, have been taken from the non-hysteretic kink observed 
in the magnetoresistance at higher temperatures between 2.5~K and 5~K.
The hysteretic kink transition at lower temperatures, on the other 
hand, has been 
extracted from the magnetic torque measurements made in the dilution 
refrigerator, denoted here by vertical crosses on the rising magnetic 
field and diagonal crosses on the falling magnetic field.

The emerging phase diagram can be described as follows: 
at temperatures above $\sim$~8~K,
$\alpha$-(BEDT-TTF)$_2$KHg(SCN)$_4$ behaves like any other organic 
metal, transforming into a DW phase at low temperatures. Given the lack of
evidence for static antiferromagnetically configured 
spins\cite{miyagawa1,pratt2} and the close similarity of our 
experimentally determined phase diagram to that predicted both by McKenzie 
\cite{mckenzie1} and Zanchi {\it et al.}\cite{zanchi1},
this is more likely to be the
CDW$_0$ phase. Certainly, the first order transition into a 
new low temperature, high magnetic phase is mostly consistent with the 
theoretical phase diagram proposed by McKenzie \cite{mckenzie1}. 
What is particularly interesting in the current work is that while the high 
magnetic field phase (CDW$_x$) is expected, 
largely on theoretical grounds, to be 
a modulated CDW phase or mixed CDW-SDW hybrid, 
its transport and magnetic properties are quite unlike those observed 
in all known DW systems
\cite{gruner1}. Rather, its physical properties more closely mimic
those of a type II superconducting phase in strong magnetic fields.
\section{conclusion}
We have shown that the magnetic behaviour of 
$\alpha$-(BEDT-TTF)$_2$KHg(SCN)$_4$ at high magnetic fields has many 
of the characteristic features of a critical state model, strongly 
resembling a type II superconductor in a 
magnetic field \cite{tinkham1,hightcconf,highBsup}. 
This critical state-like behaviour, 
together with the pronounced drop in resistivity at low temperatures, 
is clearly unlike that expected for a DW systems \cite{gruner1}, 
nor can it be connected with the QHE.

While certain physical aspects of the behaviour of 
$\alpha$-(BEDT-TTF)$_2$KHg(SCN)$_4$ at high magnetic fields are 
similar to those of a type II superconductor, 
there are, however, many reasons for questioning any 
interpretation in terms of a field-induced superconducting state. 
While the presence of a CDW at low magnetic fields 
indicates that electron-phonon interactions are important in this 
material, a CDW ground state is invariably always more stable than an 
electron-phonon mediated superconductor in a strong magnetic 
field. Both CDW's and singlet-paired superconductors are Pauli limited  
at a critical field $B_{\rm c}$. 
However, superconductors have the additional disadvantage 
of being suppressed by orbital effects at much lower magnetic fields 
than that at which the Pauli critical field is expected to occur 
\cite{tinkham1}.  
Thus, if the ground state of 
$\alpha$-(BEDT-TTF)$_2$KHg(SCN)$_4$ results from the competition 
between superconductivity and CDW's, as one might gather from the fact 
that the $M=$~NH$_4$ salt is superconducting and that 
$\alpha$-(BEDT-TTF)$_2$KHg(SCN)$_4$ becomes superconducting under 
uniaxial stress, the fact that the 
CDW has already ``won'' at $B=$~0, makes the emergence of 
superconductivity in an applied field seem somewhat remote.
Furthermore, while several models have been proposed for ``reentrant'' 
superconductivity in very strong magnetic fields 
\cite{highBsup,tesanovic1,akera1,dupuis1}, 
a common prerequisite of these models is 
that the material is already a stable superconductor at $B=$~0.

From the CDW perspective, on the other hand, neither the modulated 
CDW phase nor the CDW-SDW hybrid phase have yet been the subject of 
detailed theoretical or experimental studies 
\cite{mckenzie1,zanchi1}. We cannot exclude the possibility that the  
CDW$_x$ phase itself exists in some form of critical state, which then 
only mimics the 
inductive behaviour of a superconductor in strong magnetic fields, 
and has yet to be identified. 
With such an explanation, however, there would be some difficulty 
in explaining the 
pronounced drop in the resistivity at low temperatures 
in this material. One 
could also imagine that, within the CDW$_x$ regime, the Fr\"{o}lich 
CDW mode is no longer efficiently coupled to impurities 
\cite{gruner1} or to the lattice, 
leading to dissipationless currents as $T\rightarrow$~0. 

Clearly, there exists a challenge in this material, and 
similar materials within the same series 
({\it i.e.} the $M=$~Tl and Rb salts), to find 
alternative ways of probing the possible 
presence of non-dissipative currents (or supercurrents) above 
$B\sim$~24.2~T. The present data indicates only that the 
high magnetic field phase of
$\alpha$-(BEDT-TTF)$_2$KHg(SCN)$_4$ exhibits properties that are 
qualitatively similar to those of type II superconductors in strong 
magnetic fields. On the other hand, without the existence of a type I 
superconducting phase, it would be very difficult to experimentally 
prove the existence of type II superconductivity at such strong magnetic 
fields. What is clear from these experiments is that the high 
magnetic field phase of $\alpha$-(BEDT-TTF)$_2$KHg(SCN)$_4$  
is very different from any CDW (or SDW) observed in any other 
material, nor can any of the observations in static magnetic fields 
be considered
consistent with the quantum Hall effect.
\section{acknowledgements}
The work is supported by the Department of Energy, the National 
Science Foundation (NSF) and the State of Florida. One of us (JSB), 
acknowledges the provision of an NSF grant (DMR-99-71474), and NH 
would like to thank Albert Migliori for useful discussions. One of us 
(LB) would also
like to acknowledge the provision of a FSU visiting scientist 
scolarship. We would futher like to acknowledge Eric Palm and Timothy 
Muphy for maintaining the dilution refrigerator during the lowest 
temperature experiments.

\begin{figure}
\caption{The magnetoresistance of $\alpha$-(BEDT-TTF)$_2$KHg(SCN)$_4$ 
sample 1 at selected temperatures, with the SdH frequency being 
675~T at $\theta =$~7$^\circ$. The ``phase-inversion'' effect is 
particularly pronounced in this sample at fields above the kink 
transition field, with the inverted SdH minima having resistivity 
values $\sim$~4 
times less that the minimum resistivity at zero field. 
}
\label{magres}
\end{figure}

\begin{figure}
\caption{The resistance versus temperature both at integer and 
half-integer filling factors for two different samples (sample 1 and 
sample 2) plotted on a logarithmic scale. 
In both samples, the resistance varies weakly with temperature 
until it drops abruptly at $\sim$~3~K, falling in an exponential 
fashion. While the effect is more abrupt at integer filling factors, a 
sharp kink is also observed at half integer filling factors.
}
\label{res}
\end{figure}

\begin{figure}
\caption{An example of the magnetic torque measured in sample 3 at the 
same orientation (7$^\circ$) as that of the transport meaurements made on 
sample 1, at 500~mK, versus magnetic field squared. A quadratic fit 
has been made to the torque at fields below 15~T. Rather than
plotting $\tau$, we have plotted $-\tau$, so that the sign of the 
magnetic torque is the same as that of $\chi_\bot$.
}
\label{torquevsB}
\end{figure}

\begin{figure}
\caption{Field-cooled magnetic torque measurements made at various 
different values of magnetic field, indicating that the change in 
slope continues to be discernable even at the highest available 
static magnetic fields. The torque has been normalised by $1/B$.
}
\label{torquevsT}
\end{figure}

\begin{figure}
\caption{The magnetic torque measured for sample 3 in rising and 
falling magnetic fields between 15 and 32~T. The lower (higher) of the two 
curves corresponds to the up (down) sweep, as indicated by the lower 
(upper) arrow. The temperature of the dilution refrigerator was 27~mK.
}
\label{hystvsB}
\end{figure}

\begin{figure}
\caption{(a) An example of a hysteresis loop observed in 
$\alpha$-(BEDT-TTF)$_2$KHg(SCN)$_4$ as a result of sweeping the 
magnetic field back and forth several times over the same interval at 
different rates between 8~mTs$^{-1}$ and 50~mTs$^{-1}$. Part of the 
hysteresis obtained by sweeping the magnetic field over the entire 
range between 24.4 and 32~T is also shown. (b) The same hysteresis 
after subtracting the dHvA and monotonic background magnetic torque.
}
\label{loop}
\end{figure}

\begin{figure}
\caption{Relaxation of the magnetic torque observed at an integer filling 
factor (a) and half-integer filling factor (b), with time $t$ plotted 
on a logarithmic scale. At half-integer filling factors, the moment 
appears to decay logarithmically, as expected for a type II 
superconductor.
}
\label{relax}
\end{figure}

\begin{figure}
\caption{(a) The hysteresis observed in 
$\alpha$-(BEDT-TTF)$_2$KHg(SCN)$_4$, measured in the dilution 
refrigerator at several different temperatures, obtained by 
subtracting the down sweep data from the up sweep. The waveform of the 
hysteresis is complicated, being largest at half-integer filling 
factors at higher temperatures, then becoming more pronounced at 
integer filling factors at the lowest temperature. (b) The dHvA 
signal extracted from the same data by summing up and down sweeps. The 
monotonic background magnetic torque has been subtracted for clarity.
}
\label{hystvsT}
\end{figure}

\begin{figure}
\caption{A logarithmic plot of the maxima in the hysteresis versus 
$T$ at  half-integer filling factors ($B\sim$~30.1~T) and integer 
filling factors ($B\sim$~30.7~T).
}
\label{exphyst}
\end{figure}

\begin{figure}
\caption{A tentative phase diagram for 
$\alpha$-(BEDT-TTF)$_2$KHg(SCN)$_4$ in a magnetic field, for small 
values of $\theta$. The points have been extracted both from 
magnetotransport and magnetic torque measurements discussed in this 
work, as well as specific heat measurements of Kovalev {\it et al.}.
}
\label{diagram}
\end{figure}

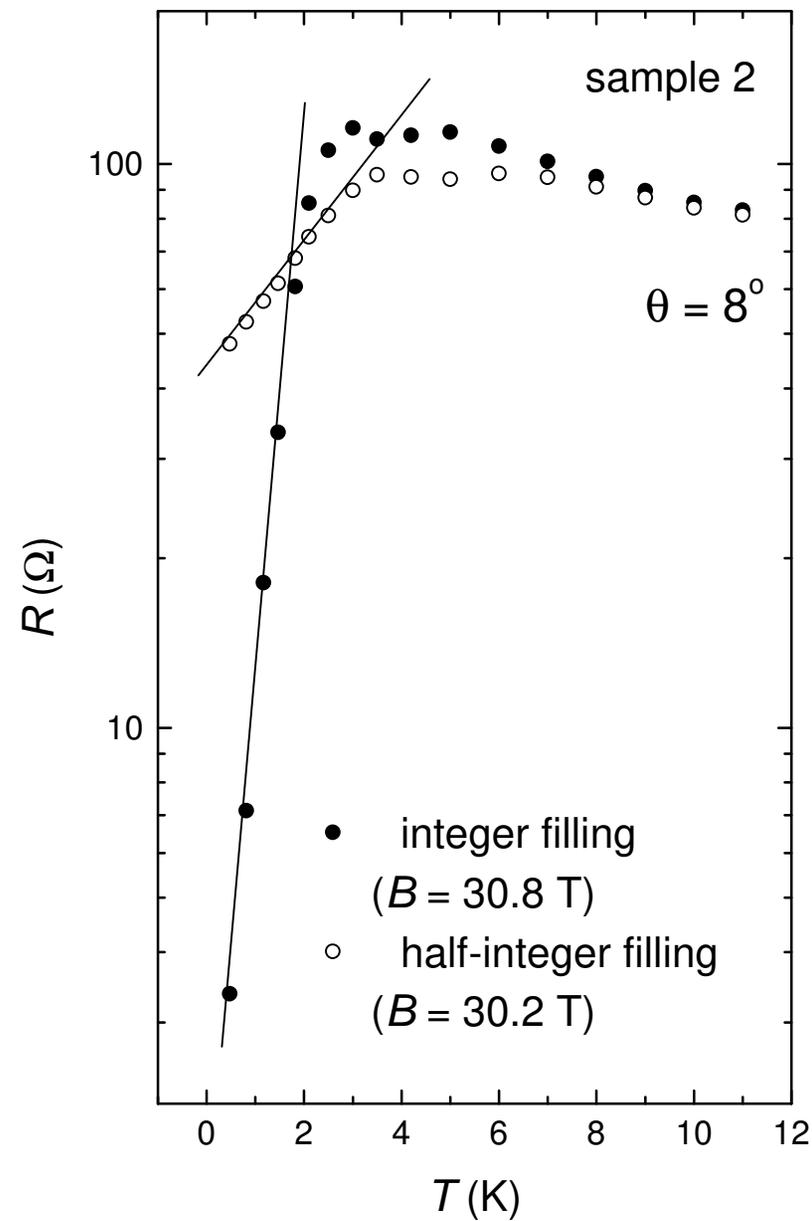
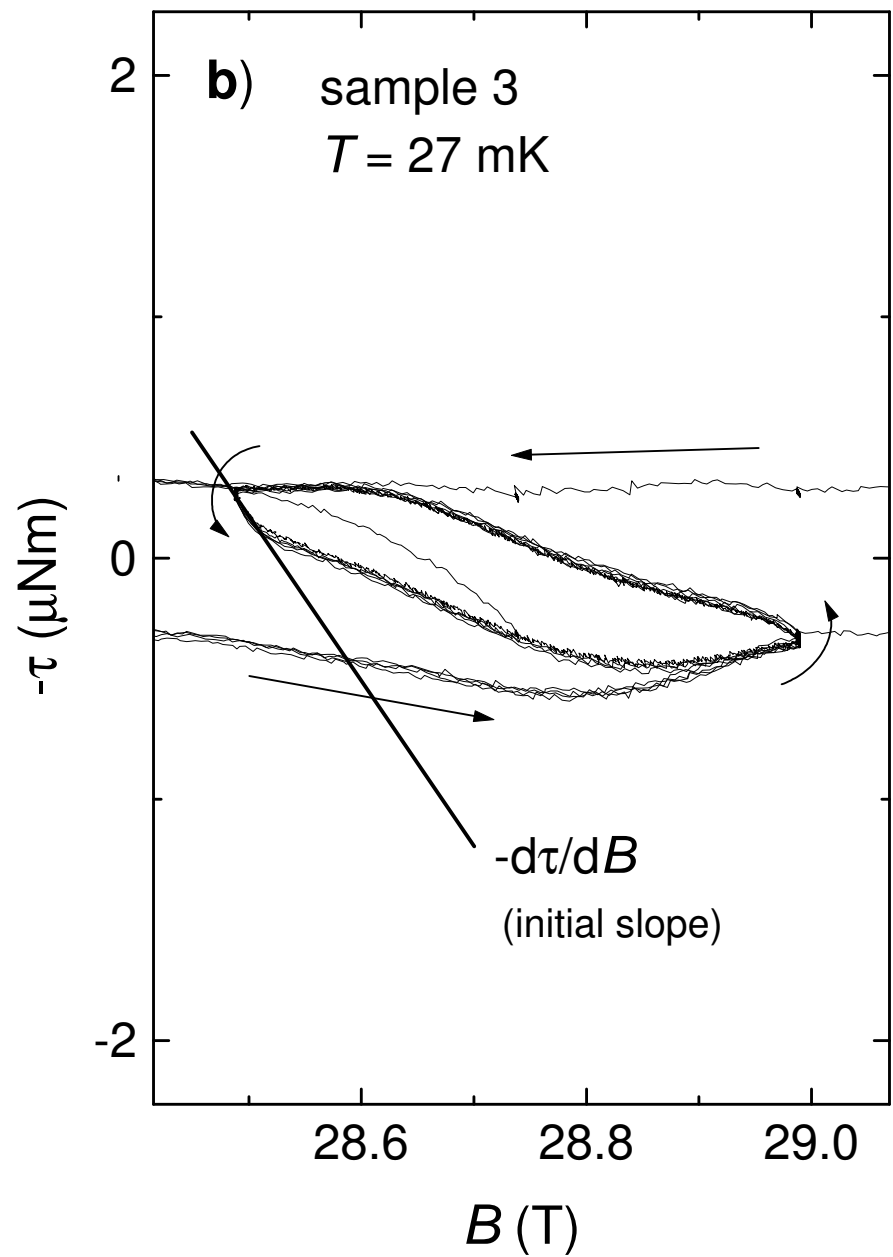
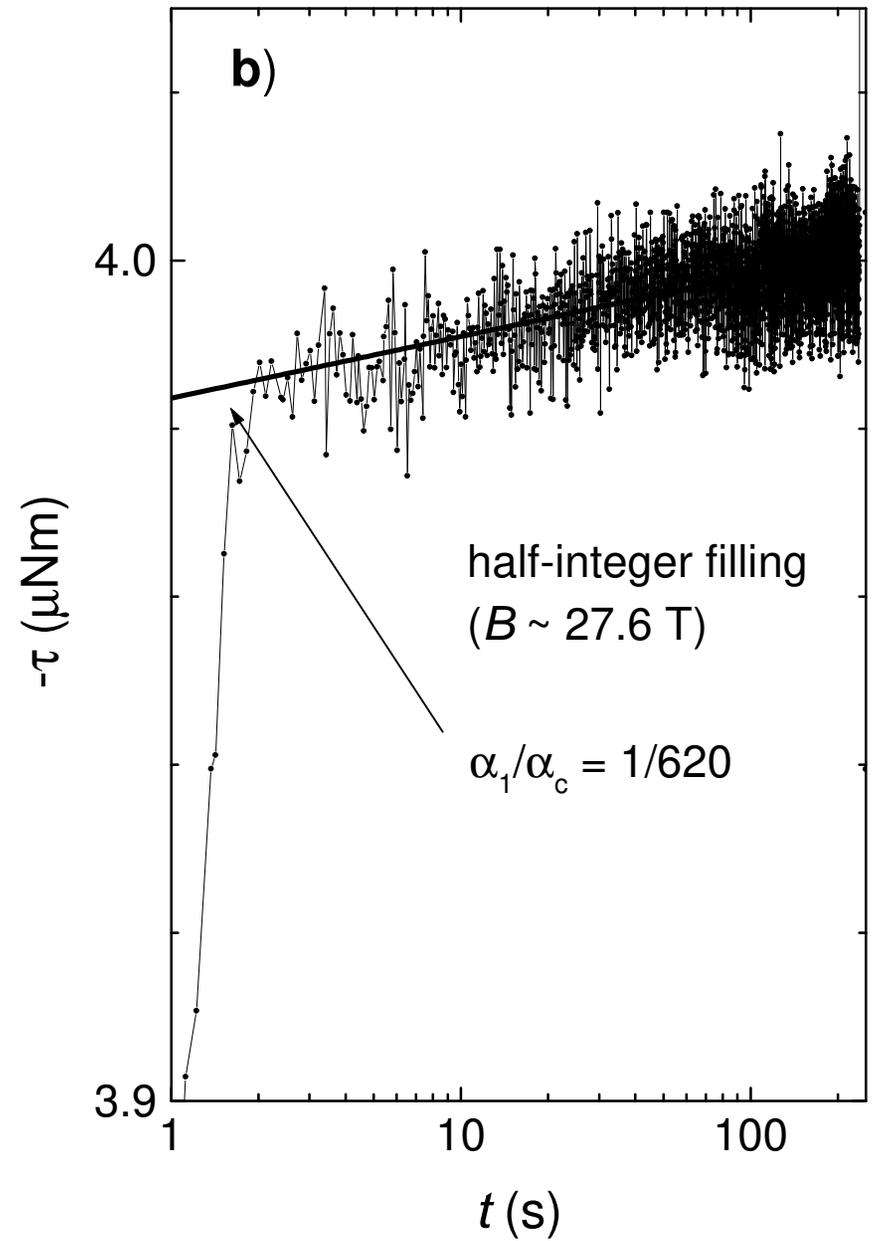
\end{document}